\documentclass[l1pt]{elsarticle}
\usepackage{lineno,hyperref}
\modulolinenumbers[5]

\journal{TCS}
\usepackage{lipsum}
\usepackage[T1]{fontenc}
\usepackage[utf8]{inputenc}
\usepackage{amsfonts, amsmath, amssymb,graphicx}
\usepackage{float}
\usepackage{parskip}
\usepackage{graphicx}
\usepackage{algorithm}
\usepackage{algpseudocode}

\newtheorem{theorem}{Theorem}

 \newcommand{\bbbn}{\mathbb{N}}

 \newcommand{\dom}{\mathrm{dom}}
 
 \newcommand{\x}{\mathbf{x}}

 \newcommand{\y}{\mathbf{y}}

\usepackage{verbatim}
\usepackage{enumerate}

\usepackage{physics}
\usepackage{times}
\usepackage{sectsty}
\usepackage{xcolor,colortbl}
\usepackage[framemethod=tikz]{mdframed}
\usepackage{caption}
\usepackage{adjustbox}
\usepackage{systeme}
\usepackage{multicol}
\begin{document}
\begin{frontmatter}
\title{How Real is Incomputability in Physics?\tnoteref{mytitlenote}}

\author{Jos\'{e} Manuel Ag\"{u}ero$^{1}$
	 Trejo, Cristian S. Calude$^{1}$,
	 Michael J. Dinneen$^{1}$,\\
	 Arkady Fedorov$^{2,3}$, Anatoly Kulikov$^{2,3}$, Rohit Navarathna$^{2,3}$, Karl Svozil$^{4}$\\[3ex]$^{1}$School of Computer Science, University of Auckland, New Zealand\\
	 $^{2}$School of Mathematics and Physics, University of Queensland, Australia\\
	 $^{3}$ARC Centre of Excellence for Engineered Quantum Systems, Queensland, Australia\\
	 $^{4}$Institut f\"{u}r Theoretische Physik, TU Wien, Vienna, Austria}
%\date{\today}
%\maketitle
\begin{abstract}
    A physical system is determined by a finite set of initial conditions and ``laws'' represented by equations. The system is computable if we can solve the equations in all instances using a ``finite body of mathematical knowledge". In this case, if the laws of the system can be coded into a computer program, then given the initial conditions of the system, one can compute the system's evolution.

%This scenario is tacitly taken for granted. But is this reasonable? The answer is negative, and a straightforward example is when the initial conditions or equations use irrational numbers, like Chaitin's Omega Number: no program can deal with such numbers because of their ``infinity''.

Are there incomputable physical systems? This question has been theoretically studied in the last 30--40 years.

 {\color{black} In this paper, we experimentally show for the first time
the strong incomputability of a quantum experiment, namely the outputs of a quantum random number generator. Moreover, the experimental results are robust and statistically significant.
}

%
%
%
%
%This article presents a class of quantum protocols producing quantum random bits. Theoretically, we prove that every infinite sequence generated by these quantum protocols is strongly incomputable -- no algorithm computing any bit of such a sequence can be proved correct. This theoretical result is not only more robust than the ones in the literature:  experimental results support and complement it.
\end{abstract}

\end{frontmatter}

\nolinenumbers

{\color{black}
 \section{Introduction}

 Incomputability in physics has been studied by many authors \cite{svozil-qct,svozil-93,svozil-pac,10.1007/978-3-642-13962-8_31,dc-d91b,computanalisysphysics,Cooper2003, incomputfoundphysics,computable_universe,DBLP:books/daglib/p/Lloyd17, Cubitt:2015aa,PhysRevX.10.031038,moore,kanter,kreisel, wolfram85b,moore,svozil-2006-ran,PhysRevLett.119.240501,abbott2012strongrandomness, Abbott:2013ly,gisin_indet,aguero_trejo_new_2021,searching_kavulich_2021}. The results in all these articles are %mainly
 theoretical, so following Einstein~\cite{EINSTEIN1936349},

\begin{quote}
Physics constitutes a logical system of thought which is in a state of evolution \dots \
%and whose basis cannot be obtained through distillation by any inductive %method from the experiences lived through, but which can only be attained %by free invention.
The justification (truth content) of the system rests in the proof of the usefulness of the resulting theorems on the basis of sense experiences, where the relations of the latter to the former can only be comprehended intuitively.
%Evolution is going on in the direction of increasing simplicity of the %logical basis. In order further to approach this goal, we must make up our %mind to accept the fact that the logical basis departs more and more from %the facts of experience and that the path of our thought from the %fundamental basis to these resulting theorems, which correlate with sense %experiences, becomes continually harder and longer.
\end{quote}

 we can ask: {\it what is their justification?} The word ``real" in the title of this article means ``a justification of incomputability based on usefulness''.

 Justifying ``usefulness'' is not easy. Indeed, for sufficiently complex systems (even reversible)
determinism on a ``one-by (to)-one'' evolution basis does not imply predictability~\cite{suppes-1993}. For example, take the $n$-body problem: the series of solutions~\cite{weierstrass-1885,poincare14,Sundman12,Wang91,Wang01} could be  ``very slowly'' convergent~\cite{Diacu96}, or even encode the Halting Problem~\cite{svozil-2007-cestial}.

In this article, we study experimentally the outputs of a quantum random number generator (QRNG), which was theoretically proven to be strongly incomputable, the only QRNG among many candidates -- see  ~\cite{Kohlrausch1926,Kragh-2009_RePoss5,qrdiodes,svozil-qct,zeilinger:qct,Quantis2020Q}. Our main results are:
a) we experimentally show
the strong incomputability of a quantum experiment, namely the outputs of a quantum random number generator, a significant improvement of the results in~\cite{abbott2018experimentally} and b)
we prove that the experimental results are robust and statistically significant.

We use a located form of the Kochen-Specker Theorem~\cite{abbott2012strongrandomness,acs-2015-info6040773,2015-AnalyticKS} to derive a class of quantum protocols producing quantum random bits~\cite{aguero_trejo_new_2021,RSPA23}. Theoretically, it was proved that every infinite sequence generated with these quantum protocols is strongly incomputable -- no algorithm computing any bit of such a sequence can be proved correct, hence the sequence is maximally unpredictable~\cite{acs-2015-info6040773}. This result is more robust than the ones in the literature and satisfies Einstein's requirement of justification: the experimental results presented here confirm and complement the theoretical results of incomputability and unpredictability and, quite significantly, the choice of physical assumptions.
}

The paper is organised as follows. In Section~\ref{3dqrngtheory}, we present the theoretical framework for the
 Localised Kochen-Specker Theorem, allowing the construction of strongly incomputable sequences via measurements of value-indefinite observables. In Section~\ref{3dqrngphysicalrealisation}, we use a standard superconducting transmon system to implement logical states as qutrits and realise the theoretical quantum protocols in Section~\ref{3dqrngtheory}.
 In Section~\ref{testing}, we present a method to empirically show the incomputability of the outputs generated in Section~\ref{3dqrngphysicalrealisation}.
 The last Section~\ref{conclusions}, we briefly discuss the results
 presented in this article and suggest further continuations.

\section{3D-QRNG -- Theory}
\label{3dqrngtheory}
%The 3D-QRNG discussed in the paper constructs a value-indefinite observable and then measures it repetitively;
%the preferred probability distribution of the outcomes is $1/4, 1/2,1/4$.

In this section, we present the theoretical framework allowing the construction of value-indefinite observables, their tolerance to measurement errors and the certification of the degree of randomness of their outcomes.

\subsection{Notation and definitions}
\label{notat}
The set of positive integers will be denoted by $\bbbn$. Consider the alphabet
$A_{b}=\{0,1,\dots, $ \\$b-1\}$, where $b\ge 2$ is an integer; the elements of
$A_b$ are the digits used in natural positional
representations of numbers in the interval $[0,1)$ at base $b$. By  $A_{b}^{*}$
and  $A_{b}^{\omega}$ we denote the sets of (finite)  strings and (infinite)
sequences over the alphabet $A_{b}$. Strings will be denoted by $x,y,u,w$; the
length of the string $x= x_1x_2\dots x_m$, $x_i\in A_{b}$, is denoted by
$|x|_{b}=m$ (the subscript $b$ will be omitted if it is clear from the
context); $A_{b}^{m}$ is the set of all strings of length $m$.  Sequences will
be denoted by $\mathbf{x}= x_1x_2\dots$; the prefix of length $m$ of
$\mathbf{x}$ is the string $ \mathbf{x}(m)= x_1x_2\dots x_m$.
Strings
will be ordered quasi-lexicographically according to the natural order $0<1<2  < \dots <b-1$ on the
alphabet $A_{b}$. For example, for $b=2$, we have $0<1<00<01<10<11<000 \dots$.
We
assume knowledge of elementary computability theory over different size
alphabets~\cite{calude:02}.

By $\mathbb{C}$, we denote the set of complex numbers.
%Sequences can be also viewed as $A_{b}$-valued functions defined on $\bbbn$.
We then fix a positive integer $n\geq 2$ and let $O \subseteq \{  P_{\psi}:  \ket{\psi} \in \mathbb{C}^{n} \}$ be a non-empty set of one-dimensional projection observables on the Hilbert space $\mathbb{C}^{n}$.

A set $C \subset O$ is a {\it  context} of $O$ if $C$ has $n$ elements  and for all $P_{\psi}, P_{\phi} \in C$ with $P_{\psi} \neq P_{\phi}, \braket{\psi}{\phi} = O$.
A  {\it  value assignment function} (on $O$) is a partial function $v:O\to \{0,1\}$ assigning values to some (possibly all) observables in $O$.  The partiality of the  function $v$ means that $v(P)$ can be $0,1$ or indefinite.
An observable $P \in O$ is  {\it  value definite} (under the assignment function $v$)  if $v(P)$ is defined, i.e.~it is 0 or 1; otherwise, it is  {\it  value indefinite} (under $v$). Similarly, we call $O$  {\it  value definite} (under $v$) if every observable $P \in O$ is value definite.

We then fix a positive integer $n\geq 2$ and let $O \subseteq \{  P_{\psi}:  \ket{\psi} \in \mathbb{C}^{n} \}$ be a non-empty set of one-dimensional projection observables on the Hilbert space $\mathbb{C}^{n}$.
A set $C \subset O$ is a {\it  context} of $O$ if $C$ has $n$ elements and for all $P_{\psi}, P_{\phi} \in C$ with $P_{\psi} \neq P_{\phi}, \braket{\psi}{\phi} = O$.
A  {\it  value assignment function} (on $O$) is a partial function $v:O\to \{0,1\}$ assigning values to some (possibly all) observables in $O$.  The partiality of the  function $v$ means that $v(P)$ can be $0,1$ or indefinite.
An observable $P \in O$ is  {\it  value definite} (under the assignment function $v$)  if $v(P)$ is defined, i.e.~it is 0 or 1; otherwise, it is  {\it  value indefinite} (under $v$). Similarly, $O$ is {\it  value definite} (under $v$) if every observable $P \in O$ is value definite.
%(a pre-existing physical property)

{\color{black}
\subsection{The quantum protocol}

The protocol is simple: {\it localise a value indefinite {\color{black}observable},  measure it, and start again afresh.}
\subsection{Localised Kochen-Specker Theorem}
%We present the main result used to construct a value indefinite observable. First,

We assume the following  premises to localise a value indefinite {\color{black}observable}.

\begin{itemize}
\item {\bf Admissibility.} This assumption guarantees agreement with quantum mechanics predictions. Fix a set $O$  of one-dimensional projection observables on $\mathbb{C}^{n}$ and the value assignment function $v:O\rightarrow \{0,1\}$. Then $v$ is  {\it  admissible} if for  every context $C$ of $O$, we have that $\sum_{P\in C}v(P) = 1$. Accordingly,  only one projection observable in a context can be assigned the value $1$.
 %$v$ is a Boolean frame function with weight $1$, so

\item {\bf Non-contextuality of definite values.}  Every outcome obtained by
measuring a value definite observable
 is {\it  non-contextual}, i.e.\ it does not
depend on other compatible
 observables, which may be measured alongside it.
 \item{\bf Eigenstate principle.}\footnote{The motivation comes from Einstein, Podolsky and Rosen's definition of {\em physical reality}~\cite[p.~777]{epr}.}  If a quantum system is prepared in the
state $\ket{\psi}$, then the projection observable
$P_\psi$ is value definite.
\end{itemize}

%The last assumption is motivated by  Einstein, Podolsky and Rosen  definition of {\em physical reality}~\cite[p.~777]{epr}:
%{\it
%         If, without in any way disturbing a system, we can predict with certainty the value of a physical quantity, then there exists a \emph{definite value} prior to observation corresponding to this physical quantity.}
%A criterion for value-definiteness results: {\it if a quantum system is prepared in an arbitrary state $\ket{\psi}\in\mathbb{C}^{n}$, then the measurement of the observable $P_{\psi}$ should yield the outcome $1$, hence, if $P_{\psi}\in O$, then $v(P_{\psi})=1.$}

%We can now state the main result:

\begin{theorem}[Localised Kochen-Specker Theorem~\cite{Abbott:2010uq,acs-2015-info6040773,PhysRevLett.119.240501, aguero_trejo_new_2021}]
\label{EffecKS}
        Assume a quantum system prepared in the state
$\ket{\psi}$ in a dimension $n\ge 3$ Hilbert space ${\mathbf C}^n$, and let $\ket{\phi}$
be any quantum state such that  $0<\abs{\bra{\psi}\ket{\phi}}<1$. If the following three conditions are satisfied: i) admissibility, ii) non-contextuality and iii)
eigenstate principle,
then the projection observable $P_\psi$ is  value
indefinite.
\end{theorem}

Theorem~\ref{EffecKS}
%states that, under the given assumptions,  any quantum state $\ket{\phi}$ that is  neither orthogonal nor parallel to $\ket{\psi}$ is {\it value indefinite}. This result
has two major consequences:

\begin{enumerate}
    \item it shows how to construct a value indefinite observable effectively,
    \item it guarantees that the status of ``value-indefiniteness'' is invariant under minor measurement errors: this is a significant property as no measurement is exact.
\end{enumerate}

We note that Theorem~\ref{EffecKS}, as the original Kochen-Specker Theorem~\cite{Kochen:2017aa}, is not valid in ${\mathbf C}^2$, hence the requirement to work in ${\mathbf C}^3$.

How ``good'' is such a 3D-QRNG, i.e.~what randomness properties can be {\it certified} for their outcomes?  For example, can we prove that the outcomes of the 3D-QRNG are ``better'' than the outcomes produced by {\it any} pseudo-random number generator (PRNG)?
For certification, we use the following assumption:
%, which is motivated by the fact that a computable sequence is the strongest form of ``deterministic hidden variable":

\begin{itemize}
        \item  {\bf epr  principle}: If a repetition of measurements of an
observable generates a computable sequence, then  these observables
are value definite.
\end{itemize}

Based on  the  Eigenstate and epr    principles, one can prove that the answer to the last question is affirmative: {\it  Any infinite repetition of the
experiment
measuring a quantum value indefinite observable generates an
incomputable infinite sequence $x_1x_2\dots$: no PRNG has this randomness property.}

A stronger result is true. Informally, a sequence $ \mathbf{x}$ is bi-immune if  no algorithm can generate
infinitely many correct values of its elements (pairs, $(i, x_i)$).
Formally, a.
 sequence $ \mathbf{x}\in A_{b}^{\omega}$  ($b\ge 2$)
 is  {\it bi-immune} if there is no
partially computable function $\varphi$ from $\bbbn $ to $A_{b}$ having an
infinite domain $\dom(\varphi)$ with the property that $\varphi(i)= x_i$ for
all $i\in \dom(\varphi)$~\cite{bienvenu2013}).

\begin{theorem}[\cite{abbott2012strongrandomness,aguero_trejo_new_2021}]
\label{biimm_2}
Assume the  Eigenstate and epr principles. An infinite repetition of the
experiment
measuring  a quantum value indefinite observable  in $\mathbb{C}^{b}$ always generates a
 $b$-bi-immune   sequence $\x\in A_2^{\omega}$, for every $b\ge 2$.
\end{theorem}

%In particular, every sequence generated by the 3D-QRNG is 3-bi-immune.

\begin{theorem}  [\cite{aguero_trejo_new_2021}]
\label{bunpredict}
Assume the epr and Eigenstate principles.
Let $\x$ be an infinite sequence obtained by measuring a quantum value indefinite observable in $\mathbb{C}^{b}$  in an infinite repetition of the experiment $E$. Then, no single bit $x_i$ can be predicted.
\end{theorem}

In particular, no single digit of every sequence   $\x \in A_3^{\omega}$ generated by the {\rm 3D-QRNG}  can be algorithmically predicted.

The following simple morphism $\varphi \colon A_3 \rightarrow A_2$ transforms a ternary
sequence into a binary sequence:

\begin{equation}
\label{amorph}
\varphi(a)=
\begin{cases}0,&\text{if }a=0,
\\1,&\text{if }a=1,
\\0&\text{if } a=2,\end{cases}
\end{equation}

\noindent, which can be extended sequentially for strings,
  $\mathbf{y}(n)=\varphi(\mathbf{x}(n))= \varphi(x_1)\varphi(x_2)$  $\dots \varphi(x_n)$ and sequences $\mathbf{y}= \varphi(\mathbf{x}) = \varphi(x_1)\varphi(x_2)\dots \varphi(x_n) \dots $. This transformation preserves 2-bimmunity:

  \begin{theorem}[\cite{aguero_trejo_new_2021}]
        Assume the epr and Eigenstate principles. Let $\mathbf{y}=\varphi (\mathbf{x})$, where $\mathbf{x}\in A_3^{\omega}$ is a ternary sequence generated by the  {\rm 3D-QRNG}  and $\varphi$ is the alphabetic morphism defined in (\ref{amorph}).
Then, no single bit of    $\y \in A_2^{\omega}$  can be predicted.
\end{theorem}

%Finally, we present the formal framework of the measurement of a valued-indefinite observable used by the 3D-QRNG.

%As noted in~\cite{aguero_trejo_new_2021},  Theorem~\ref{EffecKS} shows that given a system prepared in state $\ket{\psi}$, a one-dimensional projection observable can only be value definite if it is an eigenstate of that observable. Consequently,  for any diagonalisable observable $O$ with spectral decomposition $O=\sum_{i=1}^n\lambda_{i}P_{\lambda_{i}}$, where $\lambda_{i}$ denotes each distinct eigenvalue with corresponding eigenstate $\ket{\lambda_{i}}$,  $O$ has a predetermined measurement outcome if and only if each projector in its spectral decomposition has a predetermined measurement outcome. Thus,  the previous result holds true to the outcome of the measurement of any observable with non-degenerate spectra.
%Such generalisation is  particularly useful in the case when we use the value assignment function to represent  a  value definite observable.
These results have been used to design the following quantum operators of the 3D-QRNG.  These 3D-QRNGs operate in a succession of events of the form ``preparation,  measurement,
 reset'', iterated indefinitely many times in an algorithmic fashion~\cite{abbott2012strongrandomness}. The first 3D-QRNG was designed in~\cite{abbott2012strongrandomness}, realized in~\cite{PhysRevLett.119.240501} and analysed in~\cite{abbott2018experimentally}. While the analysis failed to observe a strong advantage of the quantum random sequences due to incomputability, it has motivated the improvement in ~\cite{aguero_trejo_new_2021}, in which the problematic probability zero branch  $S_x=0$  in Figure~\ref{fig1}.

\begin{figure}[ht]
\centering
\tikzset{every picture/.style={line width=0.60pt}} %set default line width to 0.75pt
\begin{tikzpicture}[x=0.6pt,y=0.6pt,yscale=-1,xscale=1]
\draw   (26,129) -- (149.2,129) -- (149.2,152.4) -- (26,152.4) -- cycle ;
\draw    (150,140) -- (190,139.42) ;
\draw [shift={(190,139.4)}, rotate = 539.51] [fill={rgb, 255:red, 0; green, 0; blue, 0 }  ][line width=0.75]  [draw opacity=0] (8.93,-4.29) -- (0,0) -- (8.93,4.29) -- cycle    ;
\draw   (190,129) -- (300,129) -- (300,152.4) -- (190,152.4) -- cycle ;
\draw    (300,140) -- (350,140) ;
\draw [shift={(350,140)}, rotate = 539.51] [fill={rgb, 255:red, 0; green, 0; blue, 0 }  ][line width=0.75]  [draw opacity=0] (8.93,-4.29) -- (0,0) -- (8.93,4.29) -- cycle    ;
\draw   (350,129) -- (460,129) -- (460,152.4) -- (350,152.4) -- cycle ;
\draw [dash pattern={on 0.84pt off 2.51pt}](460,140) -- (520,140) ;
\draw [shift={(520,140)}, rotate = 539.51] [fill={rgb, 255:red, 0; green, 0; blue, 0 }  ][line width=0.75]  [draw opacity=0] (8.93,-4.29) -- (0,0) -- (8.93,4.29) -- cycle    ;
\draw    (300,152.4) -- (350,183.21) ;
\draw [shift={(350,184.4)}, rotate = 216.66] [fill={rgb, 255:red, 0; green, 0; blue, 0 }  ][line width=0.75]  [draw opacity=0] (8.93,-4.29) -- (0,0) -- (8.93,4.29) -- cycle    ;
\draw    (300,129) -- (350,95.62) ;
\draw [shift={(350,94.4)}, rotate = 502.44] [fill={rgb, 255:red, 0; green, 0; blue, 0 }  ][line width=0.75]  [draw opacity=0] (8.93,-4.29) -- (0,0) -- (8.93,4.29) -- cycle    ;
\draw    (460,130) -- (520,96.62) ;
\draw [shift={(520,95.4)}, rotate = 502.44] [fill={rgb, 255:red, 0; green, 0; blue, 0 }  ][line width=0.75]  [draw opacity=0] (8.93,-4.29) -- (0,0) -- (8.93,4.29) -- cycle    ;
\draw    (460,152.4) -- (520,183.21) ;
\draw [shift={(520,184.4)}, rotate = 216.66] [fill={rgb, 255:red, 0; green, 0; blue, 0 }  ][line width=0.75]  [draw opacity=0] (8.93,-4.29) -- (0,0) -- (8.93,4.29) -- cycle    ;
\draw   (520,90) .. controls (520,82) and (527,75) .. (536,75) .. controls (544,75) and (552,82) .. (552,90) .. controls (552,99) and (544,106) .. (536,106) .. controls (527,106) and (520,99) .. (520,90) -- cycle ;
\draw   (520,186) .. controls (520,178) and (527,171) .. (536,171) .. controls (544,171) and (552,178) .. (552,186) .. controls (552,195) and (544,202) .. (536,202) .. controls (527,202) and (520,195) .. (520,186) -- cycle ;
\draw (87.6,140.7) node  [align=left] {Spin-1 source};
\draw (240.6,140.7) node  [align=left] {$\displaystyle S_{z}$ splitter };
\draw (400.6,140.7) node  [align=left] {$\displaystyle S_{x}$ splitter };
\draw (312,109) node  [align=left] {1};
\draw (312,174) node  [align=left] {\mbox{-}1};
\draw (325,131) node  [align=left] {0};
\draw (535,90.7) node  [align=left] {1};
\draw (535,185.7) node  [align=left] {0};
\draw (478,109) node  [align=left] {1};
\draw (478,174) node  [align=left] {\mbox{-}1};
\draw (492,131) node  [align=left] {0};
\draw (575,90.7) node  [align=left] {$\displaystyle\color{blue}{\frac{1}{2}}$};
\draw (575,185.7) node  [align=left] {$\displaystyle\color{blue}{\frac{1}{2}}$};
\end{tikzpicture}
\caption{QRNG setup proposed in \cite{abbott2012strongrandomness}; the values $\frac{1}{2},\frac{1}{2}$ (in blue) correspond to the outcome probabilities}
\label{fig1}
\end{figure}
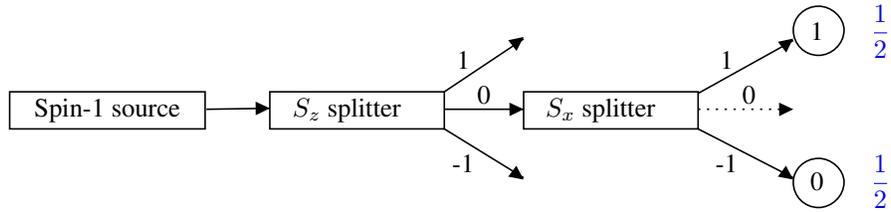

The next 3D-QRNG is presented in
Figure~\ref{fig2}. The unitary matrix $U_{x}$ corresponding to the spin state operator $S_x$ is
 $$ U_x = \frac{1}{2}
 \begin{pmatrix}
 1 & \sqrt{2} & 1\\
 \sqrt{2} & 0 & -\sqrt{2}\\
 1 & -\sqrt{2} & 1
 \end{pmatrix}.$$

As  $U_x$ can be decomposed into two-dimensional transformations~\cite{Clements:16}

$$U_x =
\begin{pmatrix}1&0&0\\
0&-i&0\\
0&0&-i
\end{pmatrix}
\cdot
\begin{pmatrix}
\frac{1}{\sqrt{3}} & \sqrt{\frac{2}{3}} & 0\\
i\sqrt{\frac{2}{3}} & -\frac{i}{\sqrt{3}} & 0\\
0&0&1\\
\end{pmatrix}
\cdot
\begin{pmatrix}
\frac{\sqrt{3}}{2} & 0 & -\frac{i}{2}\\
0&1&0\\
\frac{i}{2}& 0 & -\frac{\sqrt{3}}{2}\\
\end{pmatrix}
\cdot
\begin{pmatrix}
1&0&0\\
0&\frac{1}{\sqrt{3}} &\sqrt{\frac{2}{3}}\\
0 & i\sqrt{\frac{2}{3}} & -\frac{i}{\sqrt{3}}
\end{pmatrix}.
$$

a physical realisation of the unitary operator by a lossless beam splitter~\cite{rzbb,yurke-86} was obtained;  the new outcome probabilities are 1/4,1/2,/1/4.

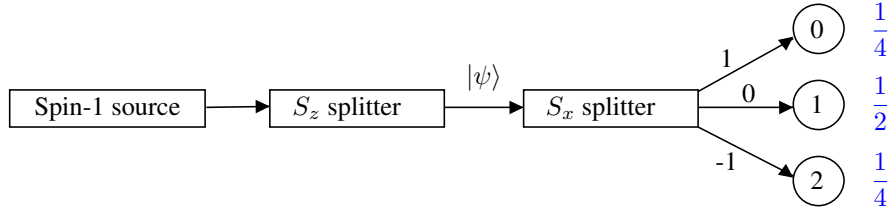
\begin{figure}
\centering
\tikzset{every picture/.style={line width=0.60pt}} %set default line width to 0.75pt
\begin{tikzpicture}[x=0.6pt,y=0.6pt,yscale=-1,xscale=1]
\draw   (26,129) -- (149.2,129) -- (149.2,152.4) -- (26,152.4) -- cycle ;
\draw    (150,140) -- (190,139.42) ;
\draw [shift={(190,139.4)}, rotate = 539.51] [fill={rgb, 255:red, 0; green, 0; blue, 0 }  ][line width=0.75]  [draw opacity=0] (8.93,-4.29) -- (0,0) -- (8.93,4.29) -- cycle    ;
\draw   (190,129) -- (300,129) -- (300,152.4) -- (190,152.4) -- cycle ;
\draw    (300,140) -- (350,140) ;
\draw [shift={(350,140)}, rotate = 539.51] [fill={rgb, 255:red, 0; green, 0; blue, 0 }  ][line width=0.75]  [draw opacity=0] (8.93,-4.29) -- (0,0) -- (8.93,4.29) -- cycle    ;
\draw   (350,129) -- (460,129) -- (460,152.4) -- (350,152.4) -- cycle ;
\draw (460,140) -- (520,140) ;
\draw [shift={(520,140)}, rotate = 539.51] [fill={rgb, 255:red, 0; green, 0; blue, 0 }  ][line width=0.75]  [draw opacity=0] (8.93,-4.29) -- (0,0) -- (8.93,4.29) -- cycle    ;
\draw    (460,130) -- (520,96.62) ;
\draw [shift={(520,95.4)}, rotate = 502.44] [fill={rgb, 255:red, 0; green, 0; blue, 0 }  ][line width=0.75]  [draw opacity=0] (8.93,-4.29) -- (0,0) -- (8.93,4.29) -- cycle    ;
\draw    (460,152.4) -- (520,183.21) ;
\draw [shift={(520,184.4)}, rotate = 216.66] [fill={rgb, 255:red, 0; green, 0; blue, 0 }  ][line width=0.75]  [draw opacity=0] (8.93,-4.29) -- (0,0) -- (8.93,4.29) -- cycle    ;
\draw   (520,90) .. controls (520,82) and (527,75) .. (536,75) .. controls (544,75) and (552,82) .. (552,90) .. controls (552,99) and (544,106) .. (536,106) .. controls (527,106) and (520,99) .. (520,90) -- cycle ;
\draw   (520,138) .. controls (520,130) and (527,123) .. (536,123) .. controls (544,123) and (552,130) .. (552,138) .. controls (552,147) and (544,154) .. (536,154) .. controls (527,154) and (520,147) .. (520,138) -- cycle ;
\draw   (520,186) .. controls (520,178) and (527,171) .. (536,171) .. controls (544,171) and (552,178) .. (552,186) .. controls (552,195) and (544,202) .. (536,202) .. controls (527,202) and (520,195) .. (520,186) -- cycle ;
\draw (87.6,140.7) node  [align=left] {Spin-1 source};
\draw (240.6,140.7) node  [align=left] {$\displaystyle S_{z}$ splitter };
\draw (400.6,140.7) node  [align=left] {$\displaystyle S_{x}$ splitter };
\draw (325,121) node  [align=left] {$\ket{\psi}$};
\draw (535,90.7) node  [align=left] {0};
\draw (535,137.7) node  [align=left] {1};
\draw (535,185.7) node  [align=left] {2};
\draw (478,109) node  [align=left] {1};
\draw (478,174) node  [align=left] {\mbox{-}1};
\draw (492,131) node  [align=left] {0};
%\draw (575,90.7) node  [align=left] {$\displaystyle\color{black}{\frac{1}{4}}$};
%\draw (575,137.7) node  [align=left] {$\displaystyle\color{black}{\frac{1}{2}}$};
%\draw (575,185.7) node  [align=left] {$\displaystyle\color{black}{\frac{1}{4}}$};
%\draw (615,90.7) node  [align=left] {$\displaystyle\color{red}{\frac{1}{3}}$};
%\draw (615,137.7) node  [align=left] {$\displaystyle\color{red}{\frac{1}{3}}$};
%\draw (615,185.7) node  [align=left] {$\displaystyle\color{red}{\frac{1}{3}}$};
\draw (575,90.7) node  [align=left] {$\displaystyle\color{blue}{\frac{1}{4}}$};
\draw (575,137.7) node  [align=left] {$\displaystyle\color{blue}{\frac{1}{2}}$};
\draw (575,185.7) node  [align=left] {$\displaystyle\color{blue}{\frac{1}{4}}$};
\end{tikzpicture}
\caption{Blueprint for a new QRNG; the values $\frac{1}{4}, \frac{1}{2},\frac{1}{4}$ (in blue)  correspond to the outcome probabilities of setups prepared in the  state $\ket{\psi} = \ket{\pm 1}$}
\label{fig2}
\end{figure}

}

\section{3D-QRNG -- Physical Realisation}

\label{3dqrngphysicalrealisation}
%\subsection{Experimental Results}
To realise the protocols shown in  Figs.~\ref{fig1},\ref{fig2} we used a standard superconducting transmon system~\cite{PhysRevLett.119.240501}. The transmon has a weakly anharmonic multi-level structure~\cite{Koch2007}, and its three lowest energy eigenstates $|0\rangle, |1\rangle$ and $|2\rangle$ can be used as the logical states of a qutrit.

To implement the protocol shown in Fig.~\ref{fig1} we followed the recipe from~\cite{PhysRevLett.119.240501} where the eigenstates of the $S_z$ operator are mapped to the states of the qutrit as follows
\begin{equation}
\{|z,-1\rangle,|z,0\rangle,|z,+1\rangle\} \rightarrow \{|2\rangle,|0\rangle,|1\rangle\}.
\end{equation}
This mapping provided an advantage of preparing $|z,0\rangle$ state by cooling down the transmon to the base temperature of a dilution refrigerator ($\sim20\,$mK).

To perform an arbitrary rotation of the qutrit quantum state $R_{\hat n}^{i,i+1}(\phi)$ we applied microwave pulses resonant to the $|0\rangle  \leftrightarrow |1\rangle$ or $|1\rangle \leftrightarrow |2\rangle$ transition frequencies, respectively. Two rotations $R^{12}_y(\pi)\cdot R^{01}_y(\pi/2)$ of the state before the dispersive measurement were used to engineer a measurement in the eigenbasis of $S_x$. The resulting measurement outcomes of the transmon energy eigenstates were mapped to the following outcomes of the measurement of $S_x$ operator: $\{|0\rangle,|1\rangle,|2\rangle\} \rightarrow \{|x,+1\rangle,|x,-1\rangle,|x,0\rangle\}$.

To implement the protocol shown in Fig.~\ref{fig2}, we used a slightly different encoding:
\begin{equation}
\{|z,-1\rangle,|z,0\rangle,|z,+1\rangle\} \rightarrow \{|1\rangle,|2\rangle,|0\rangle\}.
\end{equation}
In this case, the state $|z,+1\rangle$ was prepared by cooling the transmon. The following measurement in the eigenbasis of $S_x$ was engineered by applying the same rotations $R^{01}_y(\pi/2)\cdot R^{12}_y(\pi/2)$ before the dispersive measurements. The measurement outcomes of the transmon were then mapped to the following outcomes of the measurement of $S_x$ operator: $\{|0\rangle,|1\rangle,|2\rangle\} \rightarrow \{|x,0\rangle,|x,-1\rangle,|x,+1\rangle\}$.

To measure the transmon, we used the standard dispersive readout scheme where the transmon is capacitively coupled to a co-planar waveguide resonator. The difference between the frequency of the resonator ($f_r = 7.63$~GHz) and the $|0\rangle\leftrightarrow|1\rangle$ ($f_{01} = 5.49$~GHz) and $|1\rangle\leftrightarrow|2\rangle$ ($f_{12} = 5.16$~GHz) transitions of the transmon was designed to be much larger than the qubit-resonator coupling to ensure that the system is in the dispersive regime. In this regime, the frequency of the resonator depended on the states of the transmon and underwent shifts of $-8.5$~MHz or $-15.5$~MHz when the transmon was excited in $|1\rangle$ or $|2\rangle$ states, relative to $f_r$ when the transmon was prepared in its ground state $|0\rangle$~\cite{Koch2007}. We used a Josephson parametric amplifier to distinguish between three different transmon states with high fidelity. In addition, we set the readout pulse frequency close to the cavity frequency corresponding to the $|1\rangle$ state of the qutrit, which allowed the three possible qutrit states to be well separated on the I-Q plane for the time-integrated signal measured with the heterodyne detection scheme. The readout frequency was then fine-tuned to maximise the three-level readout fidelity. The measurement response was classified using a convolutional neural network (CNN) to increase the readout fidelity further, as described in~\cite{navarathnaNeuralNetworksOnthefly2021}.

The procedure used to generate the random numbers required an initial calibration procedure typical for circuit quantum electrodynamics setups. This involved calibration of $f_r$, $f_{01}$ and the $R^{01}_y(\pi)$ and $R^{01}_y(\pi/2)$ pulses. Two $R^{01}_y(\pi/2)$ pulses were used to fine-tune $f_{01}$ using a Ramsey measurement. The $R^{01}_y(\pi)$ and $R^{01}_y(\pi/2)$ pulses were then fine-tuned with repeated pulses. A similar procedure was followed to calibrate for $f_{12}$ and the $R^{12}_y(\pi)$ and $R^{12}_y(\pi/2)$ pulses.

After initial calibrations, we optimised the readout frequency of a single-shot readout using the Josephson parametric amplifier. The CNN is then trained for 50 cycles using 1024 measurements of the readout resonator after preparing each of the three states, $|0\rangle$,$|1\rangle$ and $|2\rangle$ as described in~\cite{navarathnaNeuralNetworksOnthefly2021}.

The procedure so far involved repeated measurements where the transmon was reset to  $|0\rangle$ state by waiting $35~\mu$s to reach thermal equilibrium (at a decay rate of $250$~kHz). We used an active reset protocol described in~\cite{Magnard2018} to increase the experiment cycle time. This involved a \textit{reset pulse} to transfer the $|2\rangle$ state population to the readout resonator and let it decay much faster (at a decay rate of $4$~MHz). An $R^{12}_y(\pi)$ pulse is then used to transfer the unwanted $|1\rangle$ state population to the $|2\rangle$ state, and the reset pulse was used again to transfer $|2\rangle$ state population to the readout resonator. The $R^{12}_y(\pi)$ ($40$~ns), reset pulse ($370$~ns), and a wait time ($50$~ns) for the readout resonator to decay were used four times in series to ensure the transmon is in the ground state, taking $1.84$~us in total. The reset time, the preparation pulses for the protocol and the measurement pulse time amounted to $3.2$~us, corresponding to a rate of $312.5$~kHz. To ensure robust generation of $100$~Gbit of random numbers we used the procedure in Section~\ref{data}.

{\color{black}
\section{Data Generation}\label{data}
The quantum random numbers have been generated using the procedure in Algorithm~\ref{alg:generation}. The algorithm involves intermittent checks of the CNN without a reset, if necessary, retraining the CNN and re-calibrating the transmon according to Algorithm~\ref{alg:calibration}. %\ref{appendix}.

\begin{algorithm}
\caption{Generation}\label{alg:generation}
\begin{algorithmic}[1]
\Procedure{RunIndex}{}
    \If{files exist}
        \State $r \gets 1+$ last \textit{random\_xxx.rbf} file number
    \Else
        \State \textbf{return} $r \gets 0$
    \EndIf
    \State \textbf{return} $r$
\EndProcedure
\State $T_{\text{rep}} \gets 40~\mu$s
\State Prepare $|0\rangle$,$|1\rangle$ and $|2\rangle$ \Comment{Cyclically for each repetition}
\State Create convolutional neural network (CNN)
\State Train CNN for 50 training cycles
\State $f \gets $ measurement accuracy \Comment{Assignment fidelity as defined in~\cite{navarathnaNeuralNetworksOnthefly2021}}
\State $c \gets 0$ \Comment{Calibration counter used to terminate}
\State $l \gets 0$ \Comment{Low $f$ counter used to calibrate}

\State $r \gets $\Call{RunIndex}{}

\While{r < 750}
    \While{$f < 0.86$}
        \If{$l > 20$}
            \If{$c > 5$}
                \State ERROR\Comment{Calibrated 5 times already. Failed}
            \EndIf
            \State \Call{Calibrate}{}
            \State $c \gets c+1$
            \State $l \gets 0$
        \EndIf
        \State $l \gets l+1$
        \State Train CNN for 20 more training cycles
        \State $f \gets $ measurement accuracy
    \EndWhile

    \State $T_{\text{rep}} \gets 3.2~\mu$s
    \State Program protocol pulses
    \State Measure $2^{26}$ repetitions
    \State Store measurements in \textit{random\_$r$.rbf}
    \State $T_{\text{rep}} \gets 40~\mu$s
\EndWhile
\end{algorithmic}
\end{algorithm}

Three types of errors could appear: initialisation errors, errors of the control pulses, and measurement errors. As the initialisation and control errors are calibrated to be kept within $<1\%$, the measurement error was the dominant error: this is due to the relaxation of the higher excited states of the qutrit to the lower energy states during the readout time. The typical assignment fidelities have been $95\%$, $88\%$, and  $78\%$ for the ground, first and second excited states, respectively.  All the fidelities have been continuously monitored during random number generation, and a drop in the value of the average assignment fidelity was used to trigger the re-calibration of the protocol (see Algorithm~\ref{alg:generation}). %\ref{appendix}).

\begin{algorithm}
\caption{Calibration}\label{alg:calibration}
\begin{algorithmic}[1]
\Procedure{Calibrate}{}\Comment{Calibrates the transmon preparation and readout}
\State $T_{\text{rep}} \gets 40~\mu$s
\State set measurement frequency to $f_r$
\State set previously calibrated settings
\State Ramsey frequency calibration for $f_{01}$
\State Calibrate $R^{01}_y(\pi)$ and $R^{01}_y(\pi/2)$ pulses
\State Ramsey frequency calibration for $f_{12}$
\State Calibrate $R^{12}_y(\pi)$ and $R^{12}_y(\pi/2)$ pulses
\State Calibrate reset pulse frequency
\State set measurement frequency to $f_r - 9$~MHz
\State Create convolutional neural network (CNN)
\State Train CNN for 50 training cycles
\EndProcedure
\end{algorithmic}
\end{algorithm}

}

\newpage
\section{Testing Incomputability}
\label{testing}

In this section, we present an empirical method to show the incomputability of the outputs generated in Section~\ref{3dqrngphysicalrealisation}.

\subsection{Why do we need testing?}
Why should we be interested in answering the above question? After all, incomputability is established
by mathematical proof, so why would we need experimental corroboration, a weaker argument? An example is a random number generator certified (by a mathematical proof) to always produce an incomputable infinite sequence of random bits. Indeed, the mathematical proof certifying incomputability is part of a mathematical model which uses certain physical assumptions; its veracity rests on those assumptions. The fact that each assumption is reasonable does not automatically guarantee that the set of assumptions is also reasonable globally. Experimental testing is essential not only for corroborating the conclusion of the proof but also for supporting the adequacy of the model.
Furthermore, thorough testing allows one to detect any issues with assumptions made in the
theoretical analysis of a device or its practical deployment.

Can we test incomputability with a statistical test, that is, with a method of statistical inference, to decide whether the data at hand sufficiently supports a particular hypothesis? The answer is {\it negative}. Intuitively, this is a consequence of the ``asymptotic'' nature of the notion of computability and its negation: {\it  finite variations do not change them.}  For example, if the sequence $x_1x_2 \dots x_n \dots$ is  computable (incomputable), then the sequences $y_1y_2\dots y_m x_1x_2 \dots x_n \dots$ and $x_kx_{k+1} \dots x_m \dots$
are also computable (incomputable) for every string $y_1y_2\dots y_m$ and positive integer $k$. For example, the Champernowne binary sequence~\cite{DGC1933}

\[0,1,00,01,10,11,000, \dots \]

\noindent obtained by concatenating all binary strings in shortlex order.\footnote{Strings are first sorted by increasing length,  and  strings of the same length are sorted into lexicographical order:
$0,1; 00,01,10,11; 000, 001, \dots 111; \dots$} This sequence is computable and  {\it normal},
i.e.~its digits are uniformly distributed: all digits are equally likely, all pairs of digits are equally likely, all triplets of digits are equally likely, and so on. Normality is a ``symptom'' of randomness, and computability is a  ``symptom'' of non-randomness. The Champernown sequence shows that these symptoms can be compatible; no statistical test can detect its computability, hence non-randomness.

Does this mean that incomputability cannot be ``experimentally tested''? Of course, no. In what follows, we will describe such a test used in assessing the quality of outputs of quantum random generators,~\cite{DBLP:journals/corr/abs-1004-1521,abbott2018experimentally}.

\subsection{Theory}

We continue with a topic apparently unrelated to the question discussed in this section: testing of primality of positive integers. Primality is considered computationally {\it easy} because there exist polynomial algorithms in the size of the input to solve it;  the first such algorithm was proposed in 2004~\cite{primeP}. However, every known primality polynomial algorithm is ``practically slow'', so probabilistic algorithms\footnote{Currently the best runs in time O$((\log \ n)^6)$.} are instead used~\cite{Stiglic2011}.\footnote{In contrast, factorisation of positive integers is ``thought'', but not proved,  to be a computationally {\it difficult} problem. Currently, one cannot factorise a positive integer of 500 decimal digits that is the product of two randomly chosen prime numbers. This fact is exploited in the RSA cryptosystem implementing public-key cryptography~\cite{Riesel2012}.}

The practical failure of polynomial primality tests motivated the search for probabilistic algorithms for primality~\cite{miller_prob_primality,rabin_prob_primality,solovay:84,solovay:118,Stiglic2011,Stiglic2011}.
To test the primality of  a positive integer $n$,  the Solovay-Strassen primality test generates the first $k$ natural numbers uniformly distributed between $1$ and $n -
1$, inclusive, and, for each $i\in\{i_1,\dots,i_k\}$ checks ``quickly'' the validity of a
predicate $W(i, n)$ based on Euler's criterion (called the Solovay-Strassen predicate).  If $W(i,
n)$ is true then ``$i$ is a witness of $n$'s compositeness''; hence $n$ is certainly not prime.
Otherwise, the test is inconclusive. In this case, the probability that $n$
is prime is greater than $1-2^{-k}$.  This result is based on the fact that {\it at
least half} the $i$'s between $1$ and $n - 1$ satisfy $W(i, n)$ if $n$ is
composite, and \emph{none} of them satisfy $W(i, n)$ if $n$ is
prime~\cite{Solovay77}.

%Chaitin and Schwartz~\cite{Chaitin78} showed that if  $c$ is a large %enough positive integer and  $s$ is a long enough
%$c$-random binary string\footnote{Recall that a string $s$ is  $c$-%random if $K(s) \ge |s|-c$; $|s|$  is the string length.}, then $n$ is prime if and only if  $Z(s,n)$ is true,
%where $Z$ is a predicate constructed directly from $O(\log n)$ conjunctions of negations of $W$ predicates.

In detail, we first define the Solovay-Strassen predicate $W(i,n)$ by
$$\left(\frac{i}{n}\right) i^{(n-1)/2} \not\equiv 1 \mod{n},$$
where $\left(\frac{i}{n}\right)$ is the Jacobi symbol\footnote{If the prime factorisation of the odd number $n$ is $p_1^{a_1}p_2^{a_2} \dots p_k^{a_k}$, then $\left(\frac{i}{n}\right) = \left(\frac{i}{p_1}\right)^{a_1} \left(\frac{i}{p_2}\right)^{a_2} \dots \left(\frac{i}{p_k}\right)^{a_k}$.} with $i\in\mathbb{N},i<n-1$.

If $i\geq 2$ and $W(i,n)$ is true, we say that $i$ is an \emph{Euler witness (E-witness)}. If $n > 3$ is an odd composite, and $W(i,n)$ is false for $i\geq 1$, we say $n$ is an \emph{Euler} pseudo-prime for the base $i$ or that $i$ is an \emph{Euler liar (E-liar)} for the Solovay-Strassen primality test. In particular, the set $L_{ss}(n)$ of \emph{E-liars} has at most $\frac{\phi(n)}{2}$ elements. Thus, the probability of sampling an \emph{E-liar} when performing the Solovay Strassen test is given by $\beta_n = |L_{ss}(n)|/(n-1)$

The size of $L_{ss}(n)$ varies for different odd composite numbers. Consider the Carmichael numbers, that is, composite positive integers $n$ satisfying the congruence $b^{n-1} \equiv 1 \pmod n$ for all integers $b$ relatively prime to $n$. The largest $\beta_n$ is found in a subset of Carmichael numbers with $\beta_n = \frac{1}{2}$.
% So, for $k$ trials, the probability that $n$ is a pseudoprime for our sampled bases less than ${2)^{-k}$.
A Carmichael number passes a Fermat primality test~\cite[Section 31.8]{introduc_algorithms} to every base relatively prime to the number, but few of them pass the Solovay-Strassen test. Increasingly
Carmichael numbers become ``rare''.\footnote{There are 1,401,644 Carmichael
numbers in the interval $[1, 10^{18}]$.}

Consider $s=s_{0} \dots s_{m-1}$ a binary string (of length $m$) and $n$ an integer greater than 2.  Let $k$ be the smallest integer such that
$(n-1)^{k+1} >  2^m -1$; we can thus rewrite the number whose binary
representation is $s$ into base $n-1$ and obtain the unique string $d_k d_{k-1}
\dots d_0$ over the alphabet $\{0,1,\dots, n-2\}$, that is,   $$\sum_{i=0}^{k}
d_i(n-1)^{i}=\sum_{t=0}^{m-1}s_t 2^t.$$  The predicate $Z(s,n)$ is defined by
\begin{equation}\label{eq:Zpredicate}
	Z(s,n)= \neg W(1+d_0, n) \wedge \dots  \wedge  \neg W(1+d_{k-1}, n),
\end{equation}

\noindent where $W$ is the Solovay-Strassen predicate.

The digits of $s$ (rewritten in base $n-1$) are used to define the Solovay Strassen predicates. If $n$ is a pseudo-prime for all the bases from $s$ used to construct these predicates, we say that $s$ is a $Z-liar$.

A string $s$ is  $c$-random if $K(s) \ge |s|-c$; $|s|$  is the string length and $K$ is the Kolmogorov complexity~\cite{calude:02}.

\begin{quotation} {\bf Chaitin-Schwartz Theorem. \cite{Chaitin78}}
{\it For all sufficiently large $c$, if $s$ is a $c$-random string of length $(l + 2c)$ and $n$ is an integer whose binary representation is $l$ bits long, then $Z(s, n)$ is true if and only if $n$ is prime.}
\end{quotation}

This result cannot be used to de-randomise\footnote{That is, to transform the probabilistic algorithm into an equivalent deterministic algorithm.} Solovay-Strassen probabilistic algorithm
because the set of $c$-random
strings is incomputable.\footnote{In fact, highly incomputable~\cite{calude:02}: no infinite set of $c$-random is computable. }
However, the result can be used to model strings from different random number generators to
test the quality of long binary strings by comparing their behaviour. In particular, we look at the number of \emph{Z-liars} found by each generator.

\subsection{Experimental analysis}

% \subsubsection{SS2}
% For each Carmichael number $n$, we repeatedly obtain a witness from the random string being tested (in the same manner as in the first test and using new bits for each Carmichael number) until the compositeness of $n$ is successfully witnessed. The test metric is the total number of bits used (for a given random string to test) to confirm the compositeness of all 16 digit Carmichael numbers. We calculate this as the sum, over all such Carmichael numbers $n$, of $\ceil{\log_2 n}$ times the number of Solovay-Strassen trials needed to witness the compositeness of $n$. (In this way, random bits that are read but then rejected
% because they give a witness larger than $n$ do not count.)

% \begin{figure}[H]
% \centering
% \begin{subfigure}
%     \centering
%     \includegraphics[width=0.48\textwidth]{Figures/ss2_boxplot_2023.pdf}
% \end{subfigure}
% \begin{subfigure}
%     \centering
%     \includegraphics[width=0.48\textwidth]{Figures/ss2C_boxplot_2023.pdf}
% \end{subfigure}
% \label{ss2box}
% \caption{Second Chaitin-Schwartz-Solovay-Strassen test: the total number of random bits required to verify the compositeness of all Carmichael numbers of at most 16 digits using. On the right we have the results of this test when taking complements of the original random strings.}
% \end{figure}

% For the non-complemented random strings we see that the QRNGs performed better than the PRNGs. Nonetheless, we can see a clear difference when repeating this test with the complemented random strings as only the qutrits retain an advantage over the PRNGs.

Standard statistical tests of randomness focus on properties of the distribution of bits or bit strings within sequences, failing to distinguish between pseudo-random number generators and quantum random number generators. To address this issue, in~\cite{abbott2018experimentally}, the ability of random strings to de-randomise the Solovay-Strassen probabilistic test of primality was used to compare the algorithmic randomness of strings generated by a QRNG and those produced by different PRNGs. Despite leading to mostly inconclusive results, the tests conducted showed some advantages offered by a 3D-QRNG against PRNGs with respect to the randomness of its outputs.

The following test, called the fourth Chaitin-Schwartz-Solovay-Strassen test (CSS4) in~\cite{abbott2018experimentally}, showed the highest potential for distinguishing between sources of random strings. Recall that the crucial fact is that the set of $c$-random
strings is (highly) {\it incomputable}.

We construct the Chaitin-Schwartz predicate $ Z(s,n)$ from (\ref{eq:Zpredicate})
and generate a pool of Solovay-Strassen predicates composed of the digits  $s$ in base $n-1$. Then, we fix $c=l-1$ where $l$ is the $l$-bit binary representation of $n$ and sample $s$ from chunks of $l(l+2c)$ bits in order to look for \emph{Z-liars} generated by a set of bases for the predicates extracted from the string $s$.

In~\cite{abbott2018experimentally}, Carmichael numbers were used in the majority of the tests. However, despite Carmichael numbers having a larger $L_{ss}(n)$, it is difficult to find \emph{Z-liars} due to the length of their binary representation. For example, for the smallest Carmichael number more than $70\times 2^{32}$ bits would need to be read to find a \emph{Z-liar} since the Solovay-Strassen test guarantees a predicate is true with a probability of at least one-half when
$n$ is a composite number. For smaller numbers we expect see to a larger number of \emph{Z-liars}. Thus, for this test, only odd composite numbers less than 50 were used for each round, and the process was repeatedly parsed through each string with an incremental bit offset.

Recently in~\cite{searching_kavulich_2021}, a similar approach was taken by applying these tests to a different set of PRNGs and two different QRNGs with a larger set of numbers; each string tested had a length of $2^{26}$. Once again, the QRNGs showed no clear advantage over the PRNGs. Moreover, the difficulty of finding \emph{Z-liars} led to a similar limitation in terms of numbers tested; \emph{Z-liars} were only observed for composites $n\leq 25$.
Still, an essential characteristic of this test was confirmed: its sensitivity to the size of the pool of unique bases extracted from the random strings. No \emph{Z-liars} were recorded when a repetitive structure generated by their sampling process was present. For this reason, we have a variation of this test was performed.

We tested two PRNGs and a QRNG: the Python3 Mersenne Twister-based generator, the hashing function SHA3, considered a ``cryptographically secure PRNG'' and the 3D-QRNG described in this paper.

Since the number of Solovay-Strassen tests increases with longer binary representations, the probability of observing a \emph{Z-liar} becomes smaller, so a large pool of unique bases was required to detect a significant number of \emph{Z-liars} \cite{abbott2018experimentally}. Thus, we prepared ten sets of strings of size $2^{32}$ for each generator and applied the shifting process described in~\cite{abbott2018experimentally} for the test.
The average number of \emph{Z-liars} over the composite numbers less than 50 was taken as the metric. Despite only detecting \emph{Z-liars} for composites up to 25, there was a noticeable difference between sources for the numbers 9 and 15. For these numbers, from our predicate construction, we have that a minimum of $40\times 2^{13}$ bits and $40\times 2^{10}$ bits are needed for a $c$-random string to have a chance of finding a \emph{Z-liar}; see Table \ref{ss4table}.

% See 2018 paper for more information.

\begin{table}
\centering
\scalebox{0.9}{
\begin{tabular}{||l c c c c c c c c c c||}
 \hline
  Composite number tested & 9 & 15 & 21 & 25 & 27 & 33 & 35 & 39 & 45 & 49 \\ [0.5ex]
  \hline
 sha3 & 265.6 & 60.3 & 0 & 0.2 & 0 & 0 & 0 & 0 & 0 & 0 \\
 \hline
 python3 & 260.1 & 58 & 0 & 0.3 & 0 & 0 & 0 & 0 & 0 & 0 \\
 \hline
 qutrits & 536.4 & 131.9 & 0 & 0.2 & 0 & 0 & 0 & 0 & 0 & 0 \\
 \hline
\end{tabular}}
\caption{Average number of \emph{Z-liars} sampled by composite number tested (over 10 strings of length $2^{32}$)}
\label{ss4table}
\end{table}

\begin{figure}[H]
    \centering
    \includegraphics[width=\textwidth]{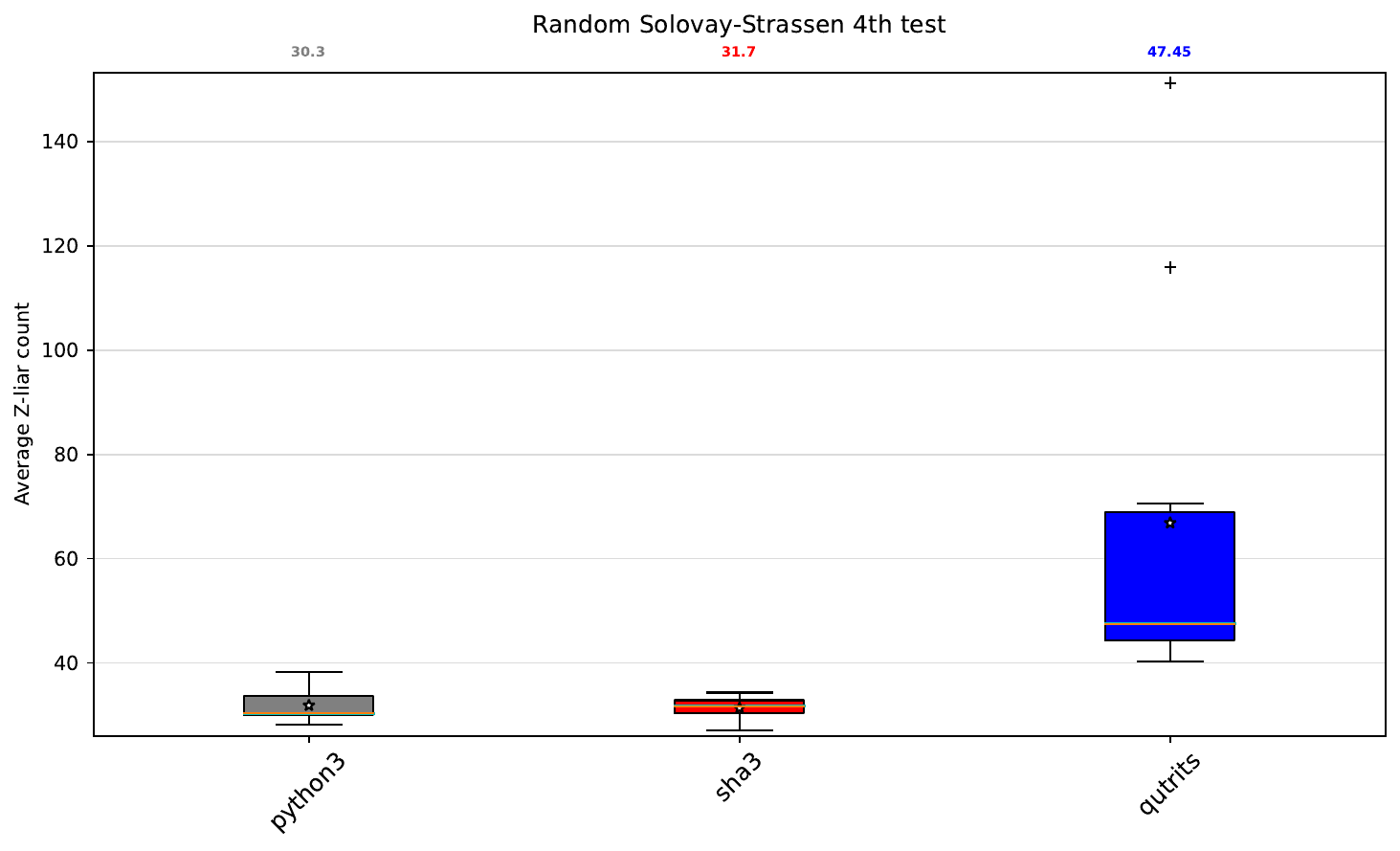}
    \caption{Fourth Chaitin-Schwartz-Solovay-Strassen test: distribution of the average \emph{Z-liar} counts for all odd composite numbers less than 50}
    \label{ss4box}
\end{figure}

The occurrence of patterns in long enough sequences of random events is inevitable. Since a lower quality of randomness increases the rate at which this occurs, the gap between the number of unique bases extractable between RNGs with different qualities of randomness widens. Thus, given long enough strings, we can observe this behaviour. Since many unique bases are required to increase the likelihood of finding \emph{Z-liars}, from Figure~\ref{ss4table}, we see the advantage offered by a 3D-QRNG generator over other alternative sources of randomness; see Fig. \ref{ss4box}.

In order to analyse the statistical significance of these results, we conducted the non-parametric and distribution-free two-sample Kolmogorov–Smirnov test. This test identifies if two datasets differ significantly without any prior assumption about an underlying distribution.
To this end, we say that the difference between two datasets is statistically significant if the $p$-value obtained through this test is less than 0.005. This critical $p$-value is chosen to reduce the chance of false positives as well as allow us to provide a direct comparison with results from \cite{abbott2018experimentally}; see Table \ref{ks-table}.

\begin{table}
\centering
\begin{tabular}{|| c c c ||}
\hline
& sha3 & qutrits  \\ [0.5ex]
\hline
python3 &  0.9780 & 0.0047\\
sha3 &  & 0.0047 \\
\hline
\end{tabular}
\caption{Kolmogorov-Smirnov test $p$-values for the fourth Chaitin-Schwartz-Solovay-Strassen test with the Z-liar count metric}
\label{ks-table}
\end{table}

We note that there is a significant difference between the 3D-QRNG qutrits and the PRNGs. A similar behaviour was revealed in ~\cite{abbott2018experimentally}, where despite the non-conclusive results of the fourth Chaitin-Schwartz-Solovay-Strassen test, the Kolmogorov-Smirnoff test showed that the difference between a 3D-QRNG and the other PRNGs is statistically relevant.  The outcomes of the fourth Chaitin-Schwartz-Solovay-Strassen test presented here show a stronger advantage of 3D-QRNGs over PRNGs.

\section{Conclusions}
\label{conclusions}
This article uses a located form of the Kochen-Specker Theorem to derive a physical realisation of a class of 3D-QRNGs by means of a superconducting transmon. The sequences produced by these 3D-QRNGs are strongly incomputable, a property that no other QRNG provides to date. Furthermore, we have used a non-statistical randomness test to probe experimentally the incomputability of its generated long strings: for the first time,  a provable advantage over the best PRNGs was found. This result has been achieved by using the Chaitin-Schwartz Theorem to probe the ``usefulness'' of generated quantum random bits,   a form of Einstein's justification.

These results highlight the real effects of incomputability in quantum systems and complement the theoretical certification via value indefiniteness of the class of QRNGs implemented. Furthermore,  the experimental results confirm and complement incomputability and, quite significantly, the choice of physical assumptions in the theoretical part.

Finally, there is a strong motivation for developing alternative tests capable of probing at algorithmic properties of randomness that better suit a wide range of applications where the quality of randomness needs to be assessed quickly or dynamically.

\bibliography{cris_s, arkady, svozil}
\bibliographystyle{abbrv}

\end{document}